\documentclass[11pt]{article}

\RequirePackage{etex}

\usepackage[margin=1in]{geometry}  	
\usepackage{amsfonts}
\usepackage{amsmath} 
\usepackage{amssymb}
\usepackage{centernot}
\usepackage{booktabs}
\usepackage{multirow}
\usepackage{algorithm}
\usepackage{algorithmicx}
\usepackage{bm} 
\usepackage{natbib}
\usepackage[onehalfspacing]{setspace} 

\usepackage{float}
\usepackage{caption}
\usepackage{graphicx}
\usepackage{adjustbox}
\usepackage{lscape}

\usepackage{adjustbox}

\usepackage{xcolor}

\usepackage{relsize}

\usepackage{bbm}
\usepackage{subfig}

\definecolor{forestgreen}{RGB}{34,139,34}
\usepackage{hyperref}
\hypersetup{
    colorlinks=true,
    linkcolor=magenta,
    filecolor=magenta, 
    citecolor=forestgreen,      
    urlcolor=blue
}

\usepackage{authblk}

\usepackage{amsthm}

\newtheorem{theorem}{Theorem}

\usepackage{xpatch}
\makeatletter
\xpatchcmd{\proof}{\@addpunct{.}}{\@addpunct{:}}{}{}
\makeatother

\makeatletter
\def\@hangfrom#1{\setbox\@tempboxa\hbox{{#1}}%
      \hangindent 0pt
      \noindent\box\@tempboxa}
\makeatother

\makeatletter
\newcommand{\vast}{\bBigg@{3}}
\newcommand{\Vast}{\bBigg@{4}}
\makeatother

\makeatletter
\newcommand*{\indep}{%
  \mathbin{%
    \mathpalette{\@indep}{}%
  }%
}
\newcommand*{\nindep}{%
  \mathbin{
    \mathpalette{\@indep}{\not}
  }%
}
\newcommand*{\@indep}[2]{%
  \sbox0{$#1\perp\m@th$}
  \sbox2{$#1=$}
  \sbox4{$#1\vcenter{}$}
  \rlap{\copy0}
  \dimen@=\dimexpr\ht2-\ht4-.2pt\relax
  \kern\dimen@
  {#2}%
  \kern\dimen@
  \copy0 
} 
\makeatother

\DeclareMathOperator{\E}{\textnormal{\mbox{E}}}
\DeclareMathOperator{\PP}{\textnormal{\mbox{P}}}

\DeclareMathOperator*{\argmin}{arg\,min}

\usepackage[utf8]{inputenc}
\DeclareUnicodeCharacter{200E}{}

\usepackage{tikz}
\usetikzlibrary{positioning,shapes.geometric}

\usepackage{setspace}
\usepackage{cite}

\usepackage[stable]{footmisc}

\usepackage{rotating}

\makeatletter
\def\@seccntformat#1{\@ifundefined{#1@cntformat}%
   {\csname the#1\endcsname\quad}  
   {\csname #1@cntformat\endcsname}
}
\let\oldappendix\appendix 
\renewcommand\appendix{%
    \oldappendix
    \newcommand{\section@cntformat}{\appendixname~\thesection\quad}
}
\makeatother

\usepackage[absolute,showboxes]{textpos}

\TPMargin{5pt}

\usepackage{thmtools, thm-restate}

\usepackage{datetime}

\usepackage{sectsty}
\subsectionfont{\noindent\textbf\itshape}
\subsubsectionfont{\normalfont\itshape}

\begin{document}

\title{\textbf{Tree-based Subgroup Discovery In Electronic Health Records: Heterogeneity of Treatment Effects for DTG-containing Therapies} \vspace*{0.3in} }

\author[1]{Jiabei Yang\footnote{Email: \texttt{jiabei\_yang@brown.edu}.}}
\author[2-3]{Ann W. Mwangi}
\author[4]{Rami Kantor}
\author[5-7]{Issa J. Dahabreh}
\author[3]{Monicah Nyambura}
\author[1]{Allison Delong}
\author[1]{Joseph W. Hogan}
\author[1]{Jon A. Steingrimsson}

\affil[1]{Department of Biostatistics, School of Public Health, Brown University, Providence, RI}
\affil[2]{Department of Mathematics, Physics \& Computing, School of Science \& Aerospace Studies, Moi University, Kenya}
\affil[3]{Academic Model Providing Access to Healthcare (AMPATH), Kenya}
\affil[4]{Division of Infectious Diseases, Warren Alpert Medical School, Brown University, Providence, RI}
\affil[5]{CAUSALab, Harvard T.H. Chan School of Public Health, Boston, MA, Boston, MA}
\affil[6]{Department of Epidemiology, Harvard T.H. Chan School of Public Health, Boston, MA}
\affil[7]{Department of Biostatistics, Harvard T.H. Chan School of Public Health, Boston, MA}

\maketitle{}

\thispagestyle{empty}

\clearpage

\thispagestyle{empty}

\vspace*{1in}

\noindent \title{\textbf{Abstract:}} The rich longitudinal individual level data available from electronic health records (EHRs) can be used to examine treatment effect heterogeneity. However, estimating treatment effects using EHR data poses several challenges, including time-varying confounding, repeated and temporally non-aligned measurements of covariates, treatment assignments and outcomes, and loss-to-follow-up due to dropout. Here, we develop the Subgroup Discovery for Longitudinal Data (SDLD) algorithm, a tree-based algorithm for discovering subgroups with heterogeneous treatment effects using longitudinal data by combining the generalized interaction tree algorithm, a general data-driven method for subgroup discovery, with longitudinal targeted maximum likelihood estimation.
We apply the algorithm to EHR data to discover subgroups of people living with human immunodeficiency virus (HIV) who are at higher risk of weight gain when receiving dolutegravir-containing antiretroviral therapies (ARTs) versus when receiving  non dolutegravir-containing ARTs.


\vspace*{0.3in}

\section{Introduction}

Dolutegravir (DTG) is an orally administered antiretroviral medication that in combination with other antiretroviral medications has been approved for the treatment of human immunodeficiency virus (HIV) infection \citep{world2018updated}. DTG-containing antiretroviral therapies (ARTs) have many advantages such as promising treatment efficacy, high medication resistance barrier and low cost compared with ARTs previously recommended by the World Health Organization (WHO) \citep{namsal2019dolutegravir, phillips2019risks, venter2019dolutegravir}. However, several randomized trials \citep{calmy2020dolutegravir, namsal2019dolutegravir, sax2020weight, venter2019dolutegravir, venter2020dolutegravir} and observational studies \citep{bourgi2020weight, bourgi2020greater, esber2022weight, surial2021weight} have found larger average weight gain among people receiving DTG-containing ARTs. Weight gains can have negative consequences among people living with HIV including increased risk of cardiovascular and metabolic diseases as well as other comorbidities \citep{achhra2016short, herrin2016weight, kim2012multimorbidity}. A few studies have explored how weight gain differs between subgroups defined by some individual characteristics \citep{bourgi2020weight, calmy2020dolutegravir, sax2020weight, venter2020dolutegravir}, but all of these studies focused on subgroups that were pre-specified by the investigators and did not attempt data-driven subgroup discovery.

Data-driven discovery of subgroups with heterogeneous treatment effects requires larger sample sizes than estimation of population average treatment effects \citep{dahabreh2016using}. Unlike randomized controlled trials, which are usually powered to detect average treatment effects in the overall population, electronic health records (EHRs) often contain rich longitudinal data collected on a large number of diverse individuals in a ``real world'' setting. This makes EHRs well suited for more fine grained treatment effect analyses such as discovering subgroups with heterogeneous treatment effects and estimating long-term effects of sustained treatments. However, treatment effect estimation using EHRs involves several challenges including: i) EHRs have repeated measures over time, and these measures are not temporally aligned; ii) potential time-varying confounding; and iii) potentially informative dropout (i.e.,~individuals that dropout of the EHRs systematically differ from those who do not dropout).

Tree-based methods stratify the covariate space into interpretable subgroups making them ideal for subgroup discovery \citep{su2008interaction}. In contrast, most other blackbox machine learning algorithms produce completely personalized predictions (rather than subgroup-specific predictions), and are therefore not applicable for subgroup discovery. Previous work on tree-based methods for discovering subgroups with differential treatment effects has mostly focused on cross-sectional randomized \citep{foster2011subgroup, seibold2016model, steingrimsson2019subgroup, su2009subgroup} or non-randomized data \citep{athey2016recursive, yang2022causal}. Additionally, previously proposed tree-based methods for treatment effect estimation with longitudinal data \citep{su2011interaction, wei2020precision} cannot handle both non-randomized time-varying treatment assignment and dropout.  

In this paper, we develop the first tree-based algorithm for discovering subgroups with heterogeneous treatment effects when comparing the effects of time-varying treatments using non-randomized longitudinal data, in the presence of potentially informative dropout (censoring). The algorithm we develop, referred to as the Subgroup Discovery for Longitudinal Data (SDLD) algorithm, extends the generalized interaction tree algorithm \citep{yang2022causal} by combining it with longitudinal targeted maximum likelihood estimators (TMLE) \citep{petersen2014targeted,  stitelman2012general, van2012targeted, van2018targeted, van2011targeted, van2006targeted}. The SDLD algorithm recursively splits the covariate space into disjoint subgroups using splitting decisions that are based on maximizing treatment effect heterogeneity as quantified by subgroup-specific TMLEs. TMLEs are doubly robust \citep{bang2005doubly, robins1994estimation}, semi-parametric efficient \citep{stitelman2012general, van2018targeted}, and guaranteed to take values in the support of the outcome. We apply the algorithm to perform the first data-driven discovery of subgroups with differential effect of DTG-containing ARTs on weight using EHRs of people living with HIV in western Kenya. 






\section{Data structure and target parameter} \label{sec: notation}

For $k \in \{0, \ldots, K\}$, let $\bm{L}_k$ be a vector of time-dependent covariates (that includes measures of the outcome $Y_k$) measured at time $k$, $A_k$ be the binary treatment indicator at time $k$, $C_k$ be an indicator whether a person drops out at time $k$, and $Y_{K+1}$ be the outcome at time $K+1$. The baseline covariates $\bm{L}_0$ can include covariates that do not vary over time and baseline values of time-varying covariates. We assume the following time ordering of the data
\[
\bm{O} = (\bm{L}_0, A_0, C_0, \bm{L}_1, A_1, C_1, \ldots, \bm{L}_K, A_K, C_K, Y_{K+1}).
\]
By the time ordering assumption, each variable has no causal effect on those measured before. For a vector $\bm{X} = (X_0, \ldots, X_K)$, define $\overline{\bm{X}}_k = (X_0, \ldots, X_k)$, $\underline{\bm{X}}_k = (X_{k+1}, \ldots, X_{K})$, and $\bm{X}$ without subscript indicates a vector from time 0 to $K$ (i.e., $\bm{X} = \overline{\bm{X}}_K$). We assume that if an individual drops out at time $k$, all quantities measured after time $k$ are not observed. That is, $C_k =1$ implies that $\underline{\bm{C}}_k$, $\underline{\bm{L}}_{k}$, $\underline{\bm{A}}_{k}$, and $Y_{K+1}$ are not observed. 

Let $\overline{\bm{Y}}_{K+1,i}^{\bm{a}, \bm{c}= \bm{0}}$ be the vector of potential outcomes \citep{rubin1974estimating, robins2000causal} under intervention on individual $i$ to set the treatment vector $\bm{A}_i$ to $\bm{a}$ and to abolish censoring (i.e.,~intervention to set $\bm{C}_i = \bm{0}$). Similarly, let $\overline{\bm{L}}_{K,i}^{\bm{a}, \bm{c} = \bm{0}}$ be the potential covariate trajectory of the $i$th individual under intervention to set the treatment vector $\bm{A}_i$ to $\bm{a}$ and to abolish censoring. 



Let $\bm{L}_0$ take values in $\bm{\mathcal{L}}_0$. Any set $w \subseteq \bm{\mathcal{L}}_0$ defines a subgroup consisting of individuals with $\bm{L}_0 \in w$. 
Define the treatment effect contrasting treatment regimes $\bm{a}_1$ and $\bm{a}_0$ within subgroup $w$ as $\delta(w) = \E[Y_{K+1}^{\bm{a}_1, \bm{c}= \bm{0}}|\bm{L}_0 \in w] - \E[Y_{K+1}^{\bm{a}_0, \bm{c}= \bm{0}}|\bm{L}_0 \in w]$. That is, $\delta(w)$ is the subgroup-specific average treatment effect contrasting the treatment regimes $\bm{a}_1$ and $\bm{a}_0$. Our objective is to find mutually exclusive and exhaustive subgroups of $\bm{\mathcal{L}}_0$ where splitting decisions are made to maximize the between subgroup treatment effect difference. 


\section{Subgroup Discovery for Longitudinal Data (SDLD) algorithm} \label{sec: LISA}


\subsection{Generalized interaction tree algorithm}

In this section, we briefly review the generalized interaction tree algorithm that has been used to discover subgroups with heterogeneous treatment effects for cross-sectional randomized trial data \citep{su2009subgroup, steingrimsson2019subgroup} and cross-sectional observational data \citep{yang2022causal}. 

The generalized interaction tree algorithm consists of three steps: initial tree building, pruning, and final tree selection. The algorithm starts by splitting the data into an initial tree building dataset and a validation dataset. Algorithm \ref{alg: git} describes the initial tree building process that runs on the initial tree building dataset where the generalized interaction tree algorithm recursively partitions the covariate space into subgroups 
until a pre-determined stopping criterion is met (e.g.,~there are too few observations in each subgroup to justify further splitting). At each attempt to partition the covariate space, the algorithm focuses on discovering subgroups that maximize differences between subgroup-specific treatment effect estimators (using some estimator, $\widehat \delta(w)$, of the treatment effect in subgroup $w$). In Section \ref{sec: TMLE} we describe three subgroup-specific treatment effect estimators appropriate for the longitudinal data structures presented in Section \ref{sec: notation}. For a given subgroup-specific treatment effect estimator $\widehat \delta(w)$, define the splitting criterion for splitting a subgroup (node) $w$ into two exclusive and exhaustive subgroups $l$ and $r$ as
    \begin{align}
     \label{eq: test_stat}
      \left(\frac{\widehat \delta(l) - \widehat \delta(r)}{\sqrt{\widehat{\text{Var}}\left[\widehat \delta(l) - \widehat \delta(r) \right]}} \right)^2.
    \end{align}
The splitting criterion \eqref{eq: test_stat} estimates a standardized difference between the treatment effects in the two subgroups $l$ and $r$. Selecting the subgroups $l$ and $r$ by maximizing the splitting criterion \eqref{eq: test_stat} results in a split that tries to find the pair of subgroups that have the largest difference in estimated treatment effects (i.e.,~maximizing estimated between subgroup treatment effect heterogeneity).



\begin{algorithm}[htbp]
\setstretch{1.35}
\begin{algorithmic}[1]
\vspace{0.5mm}
 \State Define the root node as consisting of all observations. Set the root node as the node under consideration.

\State Define $L_0^{(j)}$ as the j-th component of the baseline covariate vector. In the node under consideration, find all permissible $(L_0^{(j)}, c)$ pairs, $j = 1, \dots, J$, that split the covariate space into two groups defined by $l = \{L_0^{(j)} < c\}$ and $r = \{L_0^{(j)} \geq c\}$, where $J$ is the number of components in the baseline covariate vector. 
    
    \State Run through all possible splits in Step 2 and split the node under consideration into two new subgroups based on the split that maximizes the splitting criterion \eqref{eq: test_stat}. 
    
    \State If a pre-determined stopping criterion is met, stop the tree building process; otherwise, on every node that is not already split and has not met the stopping criterion, repeat Steps 2-4.
    
\end{algorithmic}
\caption{\label{alg: git} Initial tree building process for the SDLD algorithm}
\end{algorithm}

We refer to the result of Algorithm \ref{alg: git} as the initial tree $\widehat{\psi}_{\max}$. The initial tree $\widehat{\psi}_{\max}$ usually substantially overfits to the data and to reduce overfitting a pruning algorithm uses the initial tree building dataset to create a set of subtrees of $\widehat{\psi}_{\max}$ that are candidates for being the final tree (i.e.,~the final model). The pruning algorithm is described in detail in Appendix \ref{app: pruning}. In short, the pruning algorithm is based on the split complexity, which for a tree $\psi$ is defined as
\begin{equation}
\label{eq: sc}
G^{\lambda}(\psi) = \sum_{w \in W_\psi} G_w(\psi) - \lambda |W_\psi|,
\end{equation}
where  $W_\psi$ is the set of internal nodes, $G_i(\psi)$ is the value of the splitting criterion for internal node $w$, $|S_\psi|$ is the number of internal nodes, and $\lambda$ is a positive penalization parameter. The first term in equation \eqref{eq: sc} is a measure of the treatment effect heterogeneity of the tree and each additional split in the tree is guaranteed to increase (or at least not decrease) the first term. To offset that, the second term penalizes the size of the tree (a measure of the complexity of the model). For a fixed $\lambda$, a subtree of the initial tree is said to be an optimal subtree if it maximizes $G^{\lambda}(\psi)$. Varying $\lambda$ from $0$ to $\infty$ results in a sequence of subtrees of different sizes, $\widehat \psi_{0} = \widehat \psi_{\max}$, $\widehat \psi_1, \ldots \widehat \psi_{\widehat D}$, that all are optimal for different intervals of $\lambda$.

In the original classification and regression tree algorithm for outcome prediction \citep{breiman1984classification}, the final tree from the sequence of trees created by the pruning algorithm is selected by optimizing a cross-validated objective function. However, as the true treatment effect is unknown for all individuals (we only observe at most one of the potential outcomes $Y_{K+1,i}^{\bm{a}_1, \bm{c}= \bm{0}}$ or $Y_{K+1,i}^{\bm{a}_0, \bm{c}= \bm{0}}$), there is no observed treatment effect for each individual to compare the treatment effect predictions to in the cross-validation procedure. Hence, cross-validation cannot be applied directly to select the final tree in the generalized interaction tree algorithm. To overcome that, the generalized interaction tree algorithm selects the final tree by estimating the split complexity on the validation dataset. For a given tree $\widehat \psi_d$ ($d \in \{0, 1, \ldots, \widehat{D}\}$) and a fixed $\lambda$, we calculate the validation split complexity by recalculating the split complexity in equation \eqref{eq: sc} using the validation dataset. The final tree is selected as the tree that maximizes the validation split complexity. This relies on specifying a value of $\lambda$ and a common choice is to use a quantile (e.g.,~the $95$th) of the $\chi^2_1$ distribution \citep{steingrimsson2019subgroup, su2008interaction, yang2022causal}, which is the asymptotic distribution of the splitting criterion \eqref{eq: test_stat} when there is no treatment effect heterogeneity. 

By implementing the initial tree building, pruning, and final tree selection, the generalized interaction tree algorithm stratifies the covariate space $\mathcal{L}_0$ into mutually exclusive and exhaustive subgroups $\widehat w_1, \ldots, \widehat w_{\widehat{M}}$ (terminal nodes). 


\subsection{Subgroup-specific treatment effect estimation}
\label{sec: TMLE}

Implementation of the generalized interaction tree algorithm relies on estimating the subgroup-specific treatment effect $\delta(w) = \E[Y_{K+1}^{\bm{a}_1, \bm{c}= \bm{0}}|\bm{L}_0 \in w] - \E[Y_{K+1}^{\bm{a}_0, \bm{c}= \bm{0}}|\bm{L}_0 \in w]$. In this section, we describe how TMLE can be used to estimate $ \E[Y_{K+1}^{\bm{a}, \bm{c}= \bm{0}}|\bm{L}_0 \in w]$ for a treatment regime $\bm{a}$.

\subsubsection{Identifiability} 

The following identifiability assumptions are needed for the potential outcome mean $\E[Y_{K+1}^{\bm{a}, \bm{c}= \bm{0}}|\bm{L}_0 \in w]$ to be identifiable \citep{robins1986new, pearl1995probabilistic, bang2005doubly}:
\begin{itemize}
\item Consistency: For time $k$, $k\in \{0, \dots, K\}$, and for every individual receiving treatment $\overline{\bm{A}}_k = \overline{\bm{a}}_k$ that has not dropped out of the study, their observed outcome at time $k+1$ ($Y_{k+1}$) is equal to their potential outcome $Y^{\overline{\bm{a}}_k, \overline{\bm{c}}_k= \bm{0}}_{k+1}$ and their observed covariate trajectory is equal to their potential covariate trajectory.
    \item Sequential exchangeability: At time $k$, $k \in \{0, \dots, K\}$, the potential outcomes ($Y^{\overline{\bm{a}}_k, \overline{\bm{c}}_k}_{k+1}, \ldots, Y^{\bm{a}, \bm{c}}_{K+1})$ are independent of treatment $A_k$ and censoring status $C_{k}$ conditional on past treatment and covariate trajectories.
    
    
    \item Positivity: For time $k$, $k \in \{0, \dots, K\}$, and all $\overline{\bm{l}}_k$ such that the density $f(\overline{\bm{L}}_k = \overline{\bm{l}}_k, \overline{\bm{A}}_{k-1} = \overline{\bm{a}}_{k-1}, \overline{\bm{C}}_{k-1} = \bm{0}) >0$,
    \begin{align*}
        \Pr(A_k = a_k | \overline{\bm{L}}_k = \overline{\bm{l}}_k, \overline{\bm{A}}_{k-1} = \overline{\bm{a}}_{k-1}, \overline{\bm{C}}_{k-1} = \bm{0}) > 0,
    \end{align*}
where $\overline{\bm{A}}_{-1}$ is defined as the empty set and $\overline{\bm{C}}_{-1} = 0$.
The positivy assumption says that conditional on past treatment and covariate history there is a positive probability of continuing to follow the treatment regime of interest.
\end{itemize}
Using these assumptions, the potential outcome mean can be written as
\begin{align} \label{ipw-id}
    \E\left[ \frac{I(\bm{A} = \bm{a}, \bm{C} = \bm{0})}{P(\bm{A} = \bm{a}, \bm{C} = \bm{0} | \bm{L})}Y_{K+1}  \bigg| \bm{L}_0 \in w\right]\bigg/\E\left[ \frac{I(\bm{A} = \bm{a}, \bm{C} = \bm{0})}{P(\bm{A} = \bm{a}, \bm{C} = \bm{0} | \bm{L})} \bigg| \bm{L}_0 \in w\right]
\end{align}
or as a series of iterated conditional expectations \citep{robins1986new}
\begin{align}
\E[Y_{K+1}^{\bm{a}, \bm{c}= \bm{0}}|\bm{L}_0 \in w] = \E\left[\E[\dots \E\left[Y_{K+1}| \bm{L}, \bm{A} = \bm{a}, \bm{C}=\bm{0} ]\dots| \bm{L}_0, A_0 = a_0, C_0=0 \right] | \bm{L_0}\in w\right]. \label{g-comp-id}
\end{align}

\subsubsection{Estimation} 

Before describing the longitudinal TMLE we start by describing inverse probability weighting (IPW) \citep{robins1992recovery} and g-computation estimators \citep{robins1986new} that are sample analogs of expressions \eqref{ipw-id} and \eqref{g-comp-id}, respectively. Denote the probability of following the treatment regime of interest at time $k$, $k \in \{0\dots, K\}$, conditional on observed treatment and covariate history and not having dropped out from the study as  
\begin{align*}
    g_{A,k}(\overline{\bm{L}}_k, \overline{\bm{A}}_{k-1}; \overline{\bm{C}}_{k-1} = \bm{0}) = \Pr(A_k = a_k | \overline{\bm{L}}_{k}, \overline{\bm{A}}_{k-1},  \overline{\bm{C}}_{k-1} = \bm{0}).
\end{align*}
Similarly, denote the probability of not dropping out from the study at time $k$, $k \in \{k = 0,\dots, K\}$, conditional on observed treatment and covariate history as 
\begin{align*}
    g_{C,k}(\overline{\bm{L}}_k, \overline{\bm{A}}_k; \overline{\bm{C}}_{k-1} = \bm{0}) = 
     \Pr(C_k = 0 | \overline{\bm{L}}_{k}, \overline{\bm{A}}_{k},
     \overline{\bm{C}}_{k-1} = \bm{0}). 
\end{align*}

For simplicity of notation, denote $g_{A, k}(\overline{\bm{L}}_k, \overline{\bm{A}}_{k-1}; \overline{\bm{C}}_{k-1} = \bm{0})$ as $g_{A, k}$ and $g_{C, k}(\overline{\bm{L}}_k, \overline{\bm{A}}_{k}; \overline{\bm{C}}_{k-1} = \bm{0})$ as $g_{C, k}$. Define $g_{A,C} = \prod_{k = 0}^K g_{A, k} g_{C, k}$ and let $\widehat g_{A,C} = \prod_{k = 0}^K \widehat g_{A, k} \widehat g_{C, k}$ be an estimator of $g_{A,C}$ using a model for treatment assignment and a model for being censored at each time. Last, let $\widehat g_{A,C,i}$ be a prediction from the model $\widehat g_{A,C}$ for the $i$th individual (i.e.,~the predicted probability that the $i$th individual follows the treatment regime $\bm{a}$ and is not censored). Using the sample analog of identification result in equation \eqref{ipw-id}, the IPW estimator is given by
\begin{align*}
 \widehat \mu_{\text{IPW}}(w) = \frac{\sum_{\bm{L}_{0, i} \in w} \frac{I(\bm{A}_i = \bm{a}, \bm{C}_i = \bm{0}) Y_{K+1,i}}{\widehat{g}_{A, C, i}}}{\sum_{\bm{L}_{0,i} \in w} \frac{I(\bm{A}_i = \bm{a}, \bm{C}_i = \bm{0})}{\widehat{g}_{A,C, i} }}.
\end{align*}
If $\widehat{g}_{A,C}$ consistently estimates $g_{A,C}$, the IPW estimator is consistent.

The g-computation estimator, sometimes referred to as the iterative conditional expectation estimator \citep{wen2021parametric}, is calculated by iteratively estimating the conditional expectations in identification result \eqref{g-comp-id}, where all the estimators for the conditional expectation are restricted to data from subgroup $w$. That is, the g-computation estimator is implemented using the following steps:
\begin{enumerate}
    \item Estimate $\E[Y_{K+1}|\bm{A}, \bm{L}]$ using data from individuals that did not dropout (i.e., $\bm{C}=0$). Use the estimator to create predictions for each individual with $\overline{\bm{C}}_{K-1} = \bm{0}$ setting their treatment regime to $\bm{A} = \bm{a}$. Denote the predictions by $\widehat Q_{K+1}$.
    \item Use $\widehat Q_{K+1}$ as a pseudo-outcome and estimate the conditional expecation of $\widehat Q_{K+1}$ conditional on $\overline{\bm{A}}_{K-1}$ and $\overline{\bm{L}}_{K-1}$ among those with $\overline{\bm{C}}_{K-1} = \bm{0}$. Use the estimator to create predictions 
    $\widehat{Q}_{K}$. Repeat this iterative process until the conditional expectation that is estimated is only conditional on treatment at the baseline visit $A_0$ and the baseline covariates $\bm{L}_0$. Denote the resulting predictions by $\widehat Q_1$.
    \item The g-computation estimator is calculated as the subgroup average of the predictions $\widehat Q_1$.
\end{enumerate}
The g-computation estimator is consistent if all the  conditional expectation estimators required for its implementation are consistent.

The longitudinal TMLE is similar to the g-computation estimator described above, but it updates the conditional expectation of the outcome in each iterative conditional expectation model above using the treatment regime and dropout information, targeted to reduce bias \citep{kreif2017estimating, van2012targeted}. Specifically, in the first step of obtaining the g-computation estimator, we use an additional ``targeting'' step where 
we fit an intercept only model using $Y_{K+1}$ as the outcome with $\widehat Q_{K+1}$ as an offset and with weights $\frac{I(\bm{A} = \bm{a}, \bm{C} = \bm{0})}{\widehat{g}_{A,C}}$. We denote the resulting predictions by $\widehat Q^*_{K+1}$ and this targeting step is performed for each iterative conditional expectation. The TMLE estimator is then calculated as the subgroup average of the predictions $\widehat Q_1^*$. In contrast to the  inverse probability weighting and g-computation estimators, TMLE is doubly robust \citep{van2012targeted}. That is, it is consistent for the potential outcome mean when either i) all models for the treatment assignment and the dropout probability are correctly specified or ii) all models for the conditional expectation of the outcome are correctly specified. It is semi-parametric efficient when all the models are correctly specified \citep{van2012targeted}. TMLE is a substitution estimator, and is therefore, under mild assumptions, guaranteed to fall within the support of the outcome $Y_{K+1}$. While we focus on TMLE in the remainder of the manuscript, the IPW, g-computation or other estimators such as the doubly robust estimator \citep{bang2005doubly} could be used.



%

\subsection{Subgroup Discovery for Longitudinal Data (SDLD) algorithm} \label{subsec: LISA}

The Subgroup Discovery for Longitudinal Data (SDLD) algorithm is defined by using the longitudinal TMLE as the treatment effect estimator in the generalized interaction tree algorithm. In Appendix \ref{sec: simulation} we present simulations that show good finite sample performance of the SDLD algorithm for identifying the correct subgroups. The SDLD algorithm splits the covariate space into disjoint and exhaustive subgroups based on maximizing between subgroup treatment effect heterogeneity and within each subgroup a treatment effect estimator is calculated (terminal node estimator). The following theorem is proved in Appendix \ref{app: cstt_LISA}.

\begin{theorem}
If Assumptions B.1-B.5 in Appendix \ref{app: cstt_LISA} hold, the SDLD algorithm is $L_2$ consistent.
\end{theorem}

At each splitting decision, the SDLD algorithm discovers subgroups by searching over all possible covariate and split-point combinations and selecting the combination that maximizes between subgroup heterogeneity estimated by the splitting criterion \eqref{eq: test_stat}. As a result, the observed treatment effect heterogeneity is likely overestimated and that results in bias in the subgroup-specific treatment effect estimators. Data splitting can be used to get unbiased subgroup-specific treatment effect estimators \citep{fithian2014optimal, yang2022causal}. Data splitting separates the data into two disjoint and exhaustive parts. One part is used for fitting the SDLD algorithm and producing a list of subgroups and the other part is used for subgroup-specific treatment effect estimation and confidence interval construction. As the data used for subgroup discovery and subgroup-specific treatment effect estimation are independent, the data splitting process results in unbiased treatment effect estimators and asymptotically valid confidence intervals.





\section{Discovering subgroups with differential weight gain when on DTG-containing ARTs versus non DTG-containing ARTs} \label{sec: ampath_anlss}




DTG is an antiretroviral medication that belongs to the class of integrase strand transfer inhibitors. In combination with other medications, DTG has shown promising results in clinical trials and is globally used today \citep{kandel2015dolutegravir}. A concern with DTG-containing antiretroviral therapies (ARTs) has been the emergence of substantial weight gain as a potential side effect. A recent phase III trial conducted in South Africa \citep{venter2019dolutegravir, venter2020dolutegravir} compared two DTG-containing ARTs to standard of care over a 96-week follow-up period. Substantially higher weight gains were seen in the two groups receiving DTG-containing ARTs compared to the standard of care group. Another phase III trial conducted in Cameroon also showed larger weight gains when on DTG-containing ART compared to a low-dose efavirenz-based ART \citep{namsal2019dolutegravir, calmy2020dolutegravir}. Of particular concern is that \citet{lake2017fat} found that weight gain associated with DTG-containing ARTs has been shown to be associated with increased truncal fat, a known risk factor for adverse cardiovascular outcomes among people living with HIV.

Not all people are expected to be at the same risk for increased weight gain associated with DTG-containing ARTs. In the South African trial previously described, weight gain was greater for females and people with lower CD4 counts or higher viral loads. 
Similarly, in the Cameroon trial weight gain was more severe among women. Discovering subgroups of people for whom being on DTG-containing ARTs is more likely to cause substantial weight gain can help the development of treatment recommendations or monitoring plans tailored to a person’s risk profile.

The Academic Model Providing Access to Healthcare (AMPATH) partnership is a consortium of research institutions focusing on prevention and treatment of HIV in western Kenya \citep{tierney2007ampath}. AMPATH administers the AMPATH’s open-source electronic medical record system (AMRS), which is a large-scale electronic health record database that has information on clinical encounters, lab measurements, and demographic characteristics. We implemented the SDLD algorithm on data from AMRS to discover subgroups with differential effect of DTG-containing ARTs on weight.

We included people living with HIV who were at least 18 years old,
were initiating or on ART, and had data on or after July 1, 2016 in AMRS. A total of $88,367$ people living with HIV met these inclusion criteria. For each individual, we defined time 0 as July 1, 2016 if the person was enrolled in AMPATH before that time and as the enrollment date at AMPATH if the person was enrolled after July 1, 2016. To structure the data, we used time periods with a length of $200$ days and captured data at time $k$ within days $\left[200k, 200(k+1)\right)$, for $k\in \{0, \dots, K+1\}$ (e.g.,~we used data from days 0-199 to create $\bm{L}_0, A_0, C_0$ and days 200-399 to create $\bm{L}_1, A_1, C_1$). We focused on the first five time periods (1,000 days) as the data on people that were always on DTG-containing ARTs became sparse over time, resulting in unstable estimates over longer follow-up. We censored individuals at time $k$ if there was no clinical visit in the time periods after time $k$. As pregnancy leads to natural weight gain, we excluded 1,475 women that were pregnant at baseline and censored individuals at time $k-1$
if they were first recorded as pregnant at time $k$, $k = 1, \dots, K$. Last, we excluded one individual who had an data entry error whose ART start date was recorded before date of birth and $2,446$ individuals that had no outcome or treatment information available in any time period.
Hence, the analysis set consisted of 84,445 individuals who had a total of 1,178,016 unique clinical visits.

The outcome of interest was weight in kilograms at time 4 (i.e., $Y_4$ collected within days 800-999). Treatment at each time was defined as 1 if an individual was receiving a DTG-containing ART, and 0 if not. The two treatment regimes of interest were always on and never on DTG-containing ART. We included HIV viral load, systolic blood pressure, diastolic blood pressure, weight, height, whether the individual had active tuberculosis, was being treated for tuberculosis, was married or living with partner, and whether the individual was covered by the National Health Insurance Fund as time-dependent covariates. The vector of baseline covariates $\bm{L}_0$ included the values of time-varying covariates at time 0, gender, age when starting ART, age at time 0, time on ART at time 0, and whether the individual was enrolled on or after July 1, 2016. Further details on how we pre-processed the data and how we handled missing data and multiple clinic visits within each time period are provided in Appendix \ref{app: anlss_ampth}.

When implementing the SDLD algorithm, we applied data splitting by randomly selecting 60\% of the dataset (50,666 individuals) for tree building using the SDLD algorithm and 40\% (33,779 individuals) for subgroup-specific treatment effect estimation. In the tree building dataset, we randomly sampled 80\% (40,533 individuals) for initial tree building and used the remaining 20\% (10,133 individuals) as the validation set used for final tree selection. In the treatment effect estimation dataset, we constructed confidence intervals for treatment effects using the non-parametric bootstrap with 1,000 bootstrap samples. 

Implementing the SDLD algorithm requires estimating models for the probability of treatment assignment, the probability of dropping out of the study, and the estimation of the pseudo-outcome at each time. For all these models, we used generalized linear regression models with linear and additive main effects for all past covariate and treatment variables.

\begin{figure}[htbp]
    \centering
    \includegraphics[width = \textwidth]{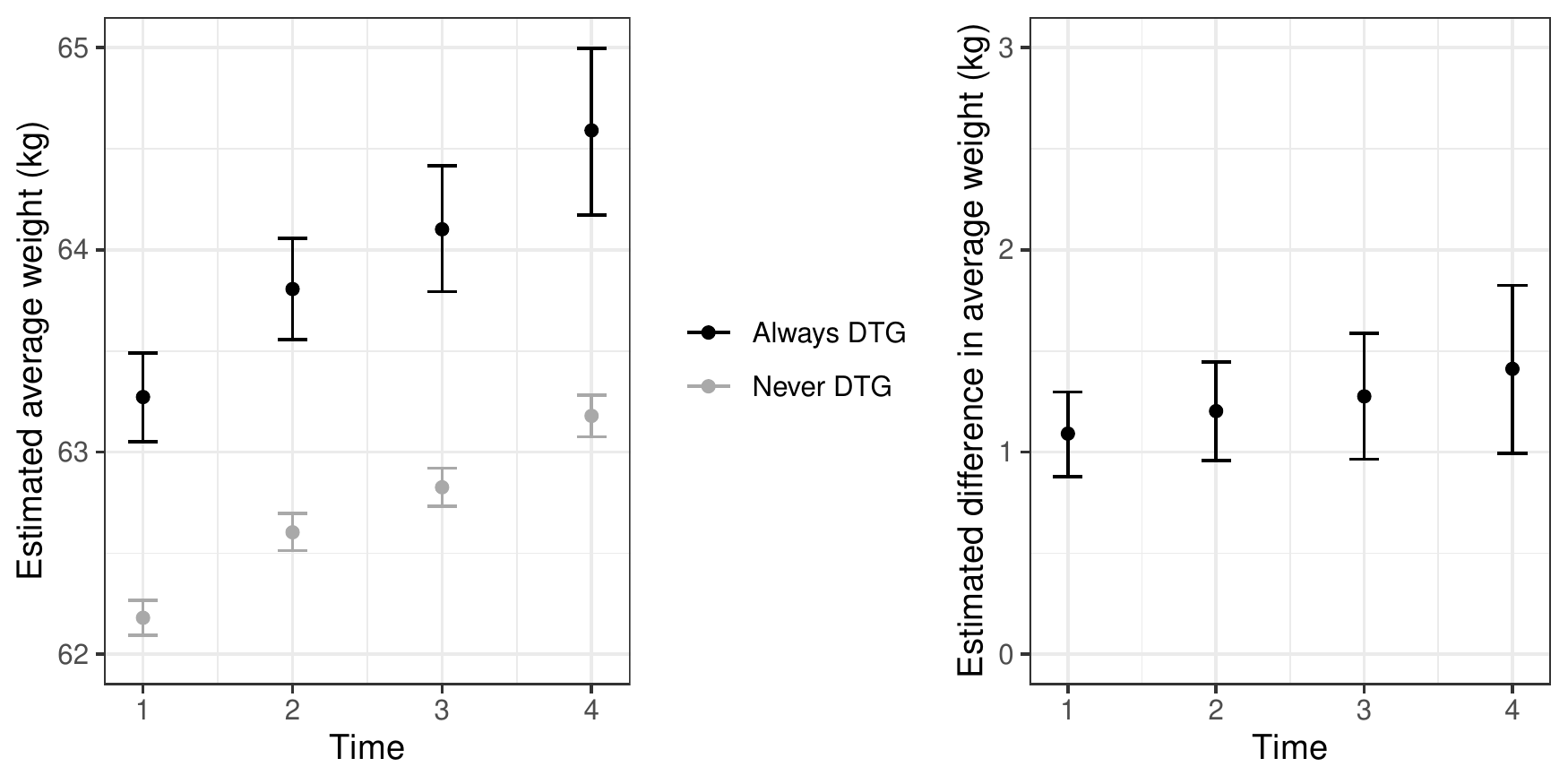}
    \caption{\label{fig: avg_eff} Estimated average weight (left) using targeted maximum likelihood estimation (TMLE) if always and never being on a DTG-containing ART and estimated causal effect on average weight (right) comparing always versus never being on a DTG-containing ART at each time. Error bars show 95\% confidence intervals estimated using the non-parametric bootstrap.}
    
\end{figure}

Figure \ref{fig: avg_eff} shows the estimated average weight had all individuals been always or never on DTG-containing ARTs (left) and the average weight gain when comparing always being on DTG-containing ARTs to never being on DTG-containing ARTs (right) at each time. The confidence intervals are estimated using the non-parametric bootstrap with 10,000 bootstrap samples. The numerical values of the estimates and the corresponding confidence intervals used for making Figure \ref{fig: avg_eff} are included in Appendix \ref{app: anlss_ampth}. The results show that sustained treatment with a DTG-containing ART leads to increased weight gain compared to never receiving DTG;  the estimated weight gain increases from 1.09 kilograms at time one (i.e., days 200-399) to 1.41 kilograms at time four (i.e., days 800-999).

\begin{figure}[htbp]
    \centering
    \includegraphics[width = \textwidth]{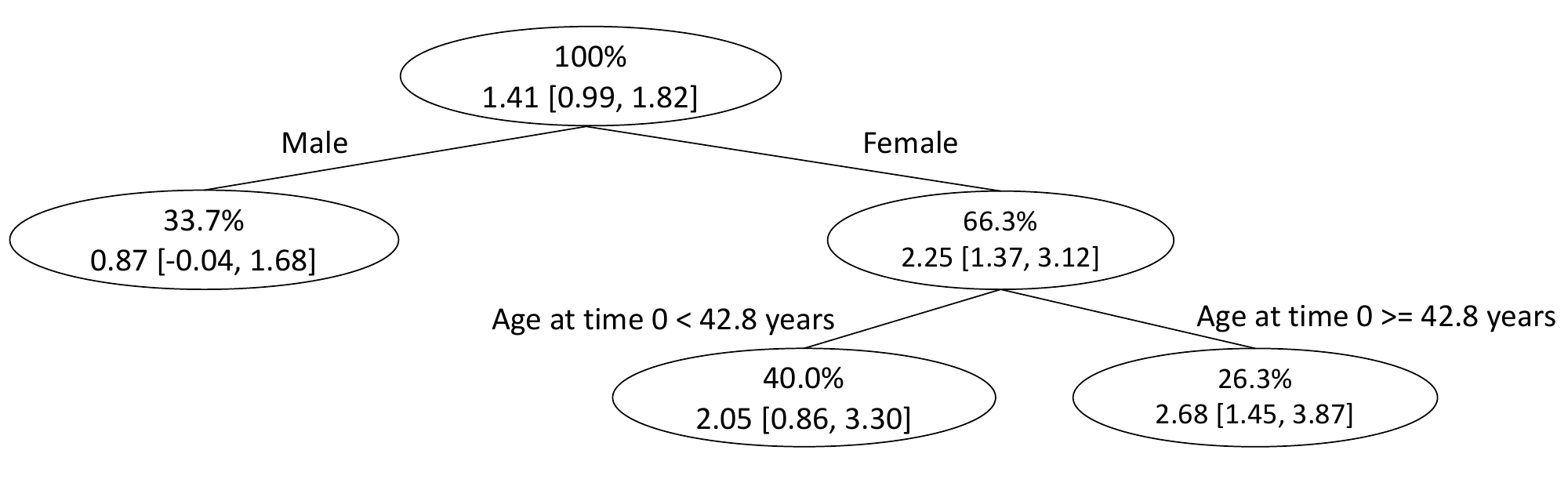}
    \caption{Final tree structure when the SDLD algorithm is applied to data from electronic health records on people living with HIV in western Kenya to discover subgroups with heterogeneous effects of DTG-containing ARTs on weight gain. Within each subgroup, the first row shows the percentage of the dataset that belongs to the subgroup and the second row shows the estimated weight gain comparing being always versus never on DTG-containing ARTs and associated 95\% confidence interval.}
    \label{fig: ampath_tree}
\end{figure}

Figure \ref{fig: ampath_tree} shows the final tree structure produced by the SDLD algorithm with subgroup-specific treatment effect estimates and associated 95\% confidence intervals. The final tree first splits on gender and then makes another split depending on whether the females are younger or older than 42.8 years old at baseline. The results suggest that the effects of DTG-containing ARTs are larger for females than males (2.25 vs 0.87 kg) although the 95\% bootstrap confidence intervals are slightly overlapping ($[-0.04, 1.68]$ for males and $[1.37, 3.12]$ for females). Results from randomized trials and other data-sources support the finding that additional weight gains due to being on DTG-containing ARTs are larger among females than males \citep{bourgi2020weight, calmy2020dolutegravir, sax2020weight, venter2020dolutegravir}. The effect of DTG-containing ARTs is estimated to be 2.05 kilograms among females that are younger than 42.8 years old and 2.68 kilograms among females that are older than or equal to 42.8 years old, but the corresponding bootstrap confidence intervals are highly overlapping ($[0.86, 3.30]$ for younger females and $[1.45, 3.87]$ for older females). In Appendix \ref{app: anlss_ampth} we present details on the estimated average weight gain from receiving DTG-containing ARTs at each time in the subgroups discovered by the SDLD algorithm and the stability of the trees produced by the SDLD algorithm.


\section{Discussion} \label{sec: dscss}

We developed the Subgroup Discovery for Longitudinal Data (SDLD) algorithm for discovering subgroups with heterogeneous treatment effects using longitudinal observational data. The algorithm combines the generalized interaction tree algorithm with longitudinal targeted maximum likelihood estimation and can handle time-varying non-randomized treatment assignment and dropout. We used the algorithm to discover subgroups of individuals with differential weight gain when receiving DTG-containing ARTs using data from AMPATH. Our results suggest that gender is the primary modifier of weight gain from receiving DTG-containing ARTs versus non DTG-containing ARTs, with females gaining more weight than males. This result is consistent with prior literature where subgroup analyses using pre-specified subgroups (not data-driven subgroup discovery) showed larger weight gains among females. 

Our analysis focused on comparing DTG-containing ARTs to other non DTG-containing ARTs and did not distinguish among different antiretroviral medications included in DTG-containing ARTs. Previous studies have evaluated different DTG-containing ARTs \citep{venter2019dolutegravir} and the results show that the weight gains might differ depending on what medications DTG is combined with. Further investigation into specific DTG-containing ARTs might provide a more fine-grained understanding of the effect of DTG on weight gain.



The SDLD algorithm defines subgroups based on baseline covariates. Extending the algorithm to dynamically update the subgroups when new information becomes available is of interest \citep{sun2020dynamic}. Although our application focuses on always and never receiving DTG-containing ARTs, the SDLD algorithm can accommodate other treatment regimes as well. Furthermore, the SDLD algorithm results in a single tree structure that creates an interpretable treatment effect stratification rule by partitioning the covariate space into identifiable subgroups with differential treatment effects. Ensemble methods, such as random forest or bagging, average multiple single trees with the aim of improving prediction accuracy. However, the resulting black-box model no longer has the structure of a single tree and unlike the single tree it does not discover clinically interpretable subgroups. It would be of interest to develop ensemble based methods for prediction with EHR data using trees built by the SDLD algorithm as the building block.

\section*{Acknowledgements}

This work was supported in part by Patient-Centered Outcomes Research Institute (PCORI) awards ME-2019C3-17875, ME-2021C2-22365; National Library of Medicine (NLM) grant R01LM013616 and
National Institute of Allergy and Infectious Diseases (NIAID) awards P30AI042853, R01AI167694 and K24AI134359. The content is solely the responsibility of the authors and does not necessarily represent the official views of the NLM, NIAID, PCORI, PCORI's Board of Governors, or PCORI's Methodology Committee.

\newpage 
\bibliography{References} 

\newpage 
\appendix 

\setcounter{page}{1}

\section{Pruning step of SDLD algorithm}
\label{app: pruning}

In this section we describe the the pruning step of the SDLD algorithm. The initial tree building step results in a large tree that usually substantially overfits the data. The goal of the pruning step is to create a sequence of subtrees of the initial tree that are candidates for being the final tree (model). This substantially reduces the computational complexity compared to looking at all possible subtrees.

The pruning algorithm we use is an extension of weakest link pruning \citep{breiman1984classification} to the setting of causal treatment effect estimation \citep{su2008interaction}. The algorithm builds a sequence of subtrees of $\widehat \psi_{\max}$ by sequentially cutting the ``weakest link'' defined as the branch of the tree with the smallest split complexity. 

For a non-terminal node $h$ define the branch that consists of the node $h$ and all its descendants as $\psi^*_h$. The split complexity of the subtree $\psi^*_h$ is zero at 
\[
\lambda^* = \frac{\sum_{w \in W_{\psi_h^*}} G_w(\psi_h^*)}{|W_{\psi_h^*}|}.
\]
If the objective is to maximize split complexity, removing the branch $\psi^*_h$ is preferred if $\lambda > \lambda^*$, and keeping branch $\psi^*_h$ is preferred if $\lambda < \lambda^*$. Weakest link pruning is given by the following steps:
\begin{enumerate}
\item Initialize the algorithm by setting $\widehat \psi_0 = \widehat \psi_{\max}$ and set $d=0$.
\item Set $d = d+1$.
\item Define a function 
\[
g(h) = 
     \begin{cases}
      \frac{\sum_{w \in W_{\psi_h^*}} G_w(\psi_h^*)}{|W_{\psi_h^*}|} &\quad\text{if $h \in W_{\widehat{\psi}_{d-1}}$} \\
       \infty & \quad\text{otherwise.} \\ 
     \end{cases}
     \]
The weakest branch of the tree is the branch rooted in node $h = \argmin_{h' \in W_{\widehat \psi_{d-1}}} g(h')$. Define the subtree $\widehat  \psi_d$ as the subtree of $\widehat \psi_{d-1}$ with all descendants of node $h$ removed.
\item Repeat steps 2 and 3 until the tree $\widehat \psi_d$ consists only of the root node.
\end{enumerate}

The weakest link pruning algorithm creates a finite sequence of trees $\widehat \psi_0, \ldots, \widehat \psi_{\widehat D}$. The final tree selection step selects a single tree from this algorithm.

\section{Consistency of SDLD algorithm} \label{app: cstt_LISA}

In this section we show that the Subgroup Discovery for Longitudinal Data (SDLD) algorithm is $L_2$ consistent. Recall that $\bm{\mathcal{L}}_0$ is the sample space of $\bm{L}_0$ and further let $\bm{\mathcal{L}}_k$ be the sample space of $\bm{L}_k$, $k \in \{1, \dots, K\}$; assume that the outcome of interest at visit $K+1$ is bounded, $|Y_{K+1}| \leq B_1 < \infty$; define the sample space of the observed data as $\mathcal{\bm{O}} = \bm{\mathcal{L}}_0\times \{0, 1\} \times \{0, 1\} \times \bm{\mathcal{L}}_1 \times \dots \times [-B_1, B_1]$.  Borrowing notation from \citet{nobel1996histogram}, let $\pi$ be any partition of $\bm{\mathcal{L}}_0$ and define $\Pi$ as the range of $\pi$; let $\hat{\psi}_n$ be an empirical rule (tree) with $\widehat{M}_n$ partitions (nodes) and define $\Pi_n$ as the range of $\hat{\psi}_n$. For each element $\pi = \{u_k\} \in \Pi$, define an associated partition $\widetilde{\pi} = \{u_k \times \{0, 1\} \times \cdots \times [-B_1, B_1]\}$ on $\mathcal{O}$ and let $\widetilde{\Pi}$ be the range of $\widetilde{\pi}$. Analogously, for each element $\pi_n = \{w_k\} \in \Pi_n$, define an associated partition $\widetilde{\pi}_n = \{w_k \times \{0, 1\} \times \cdots \times  [-B_1, B_1]\}$ on $\mathcal{O}$ and let $\widetilde{\Pi}_n$ be the range of $\widetilde{\pi}_n$. Finally, let 
$w_{\pi}(\bm{l}_0)$ be the unique partition (terminal node) in partition rule $\pi$ that contains the covariate vector $\bm{l}_0$.

Define $\bm{\theta}_{w_{\pi_n}(\bm{l_0})}^{\bm{a}}$ as a vector of the nuisance parameters used to implement TMLE for treatment regime $\bm{a}$ estimated using data from the terminal node $w_{\pi_n}(\bm{l_0})$. These nuisance parameters include those used to index the models for the probability of treatment assignment, probability of being censored, and the model for the conditional expectation of the outcome. The TMLE for $\E[Y_{K+1}^{\bm{a}, \bm{c} = \bm{0}} | \bm{L}_0 \in w_{\pi_n}(\bm{l_0})]$ is given by averaging over the predictions from all observations 
that condition on $\bm{L}_0 \in w_{\pi_n}(\bm{l_0})$. Let $Y_{i,w_{\pi_n}(\bm{l}_0)}^{*, \bm{a}}\{\bm{O}; \bm{\theta}_{w_{\pi_n}(\bm{l_0})}^{\bm{a}}\}$ be the prediction from the $i$th individual used to calculate that average. Define the transformed outcome
\[
Y_{i,w_{\pi_n}(\bm{l}_0)}^*\{\bm{O}; \bm{\theta}_{w_{\pi_n}(\bm{l_0})}\} = Y_{i,w_{\pi_n}(\bm{l}_0)}^{*, \bm{a}_1} \{\bm{O}; \bm{\theta}_{w_{\pi_n}(\bm{l_0})}^{\bm{a}_1}\} - Y_{i,w_{\pi_n}(\bm{l}_0)}^{*, \bm{a}_0} \{\bm{O}; \bm{\theta}_{w_{\pi_n}(\bm{l_0})}^{\bm{a}_0}\},
\]
where $\bm{\theta}_{w_{\pi_n}(\bm{l_0})} = (\bm{\theta}_{w_{\pi_n}(\bm{l_0})}^{\bm{a}_0}, \bm{\theta}_{w_{\pi_n}(\bm{l_0})}^{\bm{a}_1})$. For simplicity of notation, we denote the transformed outcome as $Y_{i, w_{\pi_n}(\bm{l}_0)}^* \left( \bm{\theta}_{w_{\pi_n}(\bm{l_0})}\right)$. 

Define $r(\bm{l}_0) = \E[Y_{K+1}^{\bm{a}_1, \bm{c} = \bm{0}} | \bm{L}_0 = \bm{l}_0] - \E[Y_{K+1}^{\bm{a}_0, \bm{c} = \bm{0}} | \bm{L}_0= \bm{l}_0]$ and the tree estimator for $r(\bm{l}_0)$ is given by
\begin{align*}
     &\widehat{r}_{n}(\bm{l}_0) = 
     \frac{\sum_{i=1}^n I\{\bm{L}_{0,i} \in w_{\widehat{\psi}_n}(\bm{l_0})\} Y_{i,w_{\widehat{\psi}_n}(\bm{l_0})}^*\left(\widehat{\bm{\theta}}_{w_{\widehat{\psi}_n}(\bm{l_0})}\right) }{\sum_{i=1}^n I\{\bm{L}_{0, i} \in w_{\widehat{\psi}_n}(\bm{l_0})\}}.
\end{align*}
The following theorem shows that the SDLD algorithm is $L_2$ consistent.

\begin{theorem} \label{thm: cons_LISA}
Assume that \begin{enumerate}
    \item[B.1] $Y_{K+1}$ is bounded such that $|Y_{K+1}| \leq B_1 < \infty$.
    
    \item[B.2] The number of terminal nodes $\widehat{M}_n \rightarrow \infty$ and $\widehat{M}_n = o\left\{\frac{n}{\log(n)}\right\}$.
    
    \item[B.3] The derivative $\left.\frac{\partial Y_{w_{\pi_n}(\bm{l}_0)}^*(\bm{\theta}_{w_{\pi_n}(\bm{l}_0)})}{\partial \bm{\theta}_{w_{\pi_n}(\bm{l}_0)}}\right|_{\bm{\theta}_{w_{\pi_n}(\bm{l}_0)} = \bm{\theta}^*_{w_{\pi_n}(\bm{l}_0)}}$ exists for $\pi_n \in \Pi_n$ and is uniformly bounded, where $\bm{\theta}^*_{w_{\pi_n}(\bm{l}_0)}$ is the asymptotic almost sure limit of $\widehat{\bm{\theta}}_{w_{\pi_n}(\bm{l}_0)}$.
    
    \item[B.4] For given $\bm{a}$ and $\bm{c}$, any $\widetilde{\pi} = \{\widetilde{u}_k\} \in \widetilde{\Pi}$ and $\bm{o} \in \widetilde{u}_k$, either the iterated conditional expectation of the observed outcome
    or the treatment regime mechanism and the censoring mechanism
    are correctly specified (necessary condition for consistency of TMLE).
    \item[B.5] The models $\widehat{g}_{A}$ and $\widehat{g}_{C}$ are uniformly bounded away from zero.
\end{enumerate}
If conditions $B.1$ through $B.5$ hold, the Subgroup Discovery for Longitudinal Data (SDLD) algorithm is $L_2$ consistent.
\end{theorem}

\noindent \textbf{Proof of Theorem \ref{thm: cons_LISA}:} Define 
\begin{align*}
     \widetilde{r}_{n}(\bm{l}_0) = \frac{\sum_{i=1}^n I\{\bm{L}_{0, i} \in w_{\widehat{\psi}_n}(\bm{l}_0)\} r(\bm{L}_{0, i}) }{\sum_{i=1}^n I\{\bm{L}_{0, i} \in w_{\widehat{\psi}_n}(\bm{l}_0)\}}.
\end{align*}
We can show that
\begin{align*} 
    \int_{\bm{\mathcal{L}}_0} \left|r(\bm{l}_0) - \widehat{r}_{n}(\bm{l}_0) \right|^2 \mathrm{d}\PP(\bm{l}_0) \leq 2\int_{\bm{\mathcal{L}}_0} \left|\widehat{r}_{n}(\bm{l}_0) - \widetilde{r}_{n}(\bm{l}_0) \right|^2 \mathrm{d}\PP(\bm{l}_0) + 2\int_{\bm{\mathcal{L}}_0} \left|r(\bm{l}_0) - \widetilde{r}_{n}(\bm{l}_0) \right|^2 \mathrm{d}\PP(\bm{l}_0),
\end{align*}
where $\PP(\bm{l}_0)$ is the distribution function for $\bm{l}_0$. We will show that both terms on the right hand side of the above equation are $o_P(1)$ which shows that $\widehat{r}_{n}(\bm{l}_0)$ is $L_2$ consistent.

Assumptions B.1 and B.5 ensure that there exists a $B_2>0$ such that \[
\max\left\{\bigg|Y_{w_{\pi_n}(\bm{l}_0)}^*\left(\widehat{\bm{\theta}}_{w_{\pi_n}(\bm{l}_0)}\right)\bigg|, \bigg|Y_{w_{\pi_n}(\bm{l}_0)}^*\left(\bm{\theta}^*_{w_{\pi_n}(\bm{l}_0)}\right)\bigg|\right\}\\ \leq B_2 < \infty
\]
for $\bm{O} \in \mathcal{O}$ and $\pi_n \in \Pi_n$, where $\bm{\theta}^*_{w_{\pi_n}(\bm{l}_0)}$ is the almost sure asymptotic limit of $\widehat{\bm{\theta}}_{w_{\pi_n}(\bm{L}_0)}$. Define $B^* = \max(B_1, B_2)$, for any subgroup $w$, $\gamma_w = \Pr(\bm{L}_{0} \in w)$, $n(w) = \sum_{i=1}^n I\{\bm{L}_{0,i} \in w\}$, $\widehat \gamma_w = n_w/n$, $\widehat{\bm{\theta}}_w$ as the nuisance parameters estimated using data from subgroup $w$, and $Y_{w, i}^*(\widehat{\bm{\theta}}_w)$ as the associated transformed outcome. Assumption B.1 and equation (18) in \citet{nobel1996histogram} imply
\begin{align*}
    &\int_{\bm{\mathcal{L}}_0} \left|\widehat{r}_{n}(\bm{l}_0) - \widetilde{r}_{n}(\bm{l}_0) \right|^2 \mathrm{d}\PP(\bm{l}_0) \\
    \leq &2B^* \sum_{w \in \widehat{\psi}_n} \gamma_w \left|\frac{1}{n(w)}\sum_{i=1}^n I\{\bm{L}_{0,i} \in w\} \{Y_{w, i}^*(\widehat{\bm{\theta}}_w) - r(\bm{L}_{0,i}) \}\right| \\
    \leq& 2B^* \sum_{w \in \widehat{\psi}_n}  \widehat \gamma_w \left|\frac{1}{n(w)}\sum_{i=1}^n I\{\bm{L}_{0, i} \in w\} \{Y_{w, i}^*(\widehat{\bm{\theta}}_w) - r(\bm{L}_{0, i}) \}\right|\\ 
    &+ 2B^* \sum_{w \in \widehat{\psi}_n}  \left|\frac{1}{n(w)}\sum_{i=1}^n I\{\bm{L}_{0, i} \in w\} \{Y_{w, i}^*(\widehat{\bm{\theta}}_w) - r(\bm{L}_{0, i}) \}\right| \left|\widehat \gamma_w - \gamma_w\right|\\
    \leq& 2B^* \sup_{\pi_n \in \Pi_n}\sum_{w \in \pi_n} \widehat \gamma_w \left|\frac{1}{n(w)}\sum_{i=1}^n I\{\bm{L}_{0, i} \in w\} \{Y_{w, i}^*(\widehat{\bm{\theta}}_w) - r(\bm{L}_{0, i}) \}\right| \\
    &+ 4B^{*2} \sup_{\pi_n \in \Pi_n}\sum_{w \in \pi_n} \left|\widehat \gamma_w - \gamma_w\right|.
\end{align*}
Applying Proposition 2, Lemma 3, and Theorem 7 in \citet{nobel1996histogram} with assumption B.2 gives\\ $\sup_{\pi_n \in \Pi_n}\sum_{w \in \pi_n} \left|\widehat \gamma_w - \gamma_w\right| = o_P(1)$. Hence,
\begin{align*}
    &\int_{\bm{\mathcal{L}}_0} \left|\widehat{r}_{n} (\bm{l}_0) - \widetilde{r}_{n}(\bm{l}_0) \right|^2 \mathrm{d}\PP(\bm{l}_0) \leq 2B_1^* \sup_{\pi_n \in \Pi_n}\sum_{w \in \pi_n} \left|\frac{1}{n}\sum_{i=1}^n I\{\bm{L}_{0, i} \in w\} \{Y_{w, i}^*(\widehat{\bm{\theta}}_w) - r(\bm{L}_{0, i}) \}\right| + o_P(1).
\end{align*}
Using Taylor series arguments and assumption B.3 gives
\begin{align*}
    &\int_{\bm{\mathcal{L}}_0} \left|\widehat{r}_{n}(\bm{l}_0) - \widetilde{r}_{n}(\bm{l}_0) \right|^2 \mathrm{d}\PP(\bm{l}_0) \\
    \leq& 2B^* \sup_{\pi_n \in \Pi_n}\sum_{w \in \pi_n} \left|\frac{1}{n}\sum_{i=1}^n I(\bm{L}_{0, i} \in w) \bigg\{Y_{w, i}^*(\bm{\theta}^*_w) + (\widehat{\bm{\theta}}_w - \bm{\theta}^*_w) \left.\frac{\partial Y_{w, i}^*(\bm{\theta})}{\partial \bm{\theta}}\right|_{\bm{\theta} = \bm{\theta}^*_w} - r(\bm{L}_{0, i}) \bigg\}\right| + o_P(1)\\
    \leq & 2B^* \sup_{\pi_n \in \Pi_n}\sum_{w \in \pi_n} \left|\frac{1}{n}\sum_{i=1}^n I(\bm{L}_{0, i} \in w) \{Y_{w, i}^*(\bm{\theta}^*_w) - r(\bm{L}_{0, i}) \}\right| + o_P(1),
\end{align*}
where the last equality follows as $\bm{\theta}^*_w$ is defined as the asymptotic limit of $\widehat{\bm{\theta}}_w$.

The same steps as in the proof of Theorem A.6.4 in \citet{yang2022causal} give
\[
2B^* \sup_{\pi_n \in \Pi_n}\sum_{w \in \pi_n} \left|\frac{1}{n}\sum_{i=1}^n I(\bm{L}_{0, i} \in w) \{Y_{w, i}^*(\bm{\theta}^*_w) - r(\bm{L}_{0, i}) \}\right| = o_P(1).
\]
Combining the above gives
\begin{equation*}    
\int_{\bm{\mathcal{L}}_0} \left|\widehat{r}_{n}(\bm{l}_{0}) - \widetilde{r}_{n}(\bm{l}_{0}) \right|^2 \mathrm{d}\PP(\bm{l}_{0})
    = o_P(1).
\end{equation*}

The Proof of Theorem 1 in \citet{nobel1996histogram} shows that
\[
\int_\mathcal{X} \left|r(\bm{l}_{0}) - \widetilde{r}_{n}(\bm{l}_{0}) \right|^2 \mathrm{d}\PP(\bm{l}_{0}) =o_P(1).
\]
Combining all of the above gives the desired result
\[
\int_{\bm{\mathcal{L}}_0} \left|r(\bm{l}_{0}) - \widehat{r}_{n}(\bm{l}_{0}) \right|^2 \mathrm{d}\PP(\bm{l}_{0}) = o_P(1).
\]

\section{Simulations} \label{sec: simulation}

We conducted simulations to evaluate the finite sample performance of the SDLD algorithm. 

\subsection{Data generation} 

The baseline covariate vector $\bm{L}_0$ was simulated from a five-dimensional mean zero multivariate normal distribution where diagonal elements of the variance matrix were all equal to 1 and all off-diagonal elements were equal to 0.2. The baseline treatment indicator $A_0$ was simulated from a Bernoulli distribution with parameter
\begin{align*}
    \Pr \left(A_0 = 1| \bm{L}_0\right) = \text{expit}\left[-0.5 + 0.2L_0^{(1)} + 0.2L_0^{(2)} + 0.4L_0^{(3)} + 0.5L_0^{(4)}\right],
\end{align*}
where $\text{expit}(x) = \exp(x)/(1+\exp(x))$. The baseline censoring indicator $C_0^{a_0}$ was simulated from a Bernoulli distribution with parameter
\begin{align*}
    \Pr\left(C^{a_0}_0=1| \bm{L}_0\right) = \text{expit}\left[-4 + 0.8a_0 + 0.3L_0^{(1)} - 0.3L_0^{(2)} - 0.3L_0^{(3)} + 0.1L_0^{(4)}\right].
\end{align*}
The covariate vector at time one had three elements $\bm{L}_1 = \left(Y_1, L_1^{(1)}, L_1^{(2)}\right)$. All of $Y_1^{a_0, c_0 = 0}$, $L_1^{(1), a_0, c_0 = 0}$ and $L_1^{(2), a_0, c_0 = 0}$ are normally distributed with means
\begin{align*}
    \E\left[Y_1^{a_0, c_0 = 0} | \bm{L}_0\right] =&  -3 + 0.1a_0 + 0.3L_0^{(1)} - 2a_0 I\left[L_0^{(2)} > 0.5\right] + 2L_0^{(4)} + 2L_0^{(5)}, \\
    \E\left[L_1^{(1), a_0, c_0 = 0} | \bm{L}_0\right] =& 0.2a_0 + 0.5L_0^{(1)} - 0.4L_0^{(2)} - 0.4L_0^{(3)} + 0.5L_0^{(4)} - 0.5L_0^{(5)}, \\
    \E\left[L_1^{(2), a_0, c_0 = 0} | \bm{L}_0, L_1^{(1), a_0, c_0 = 0}\right] =& 0.1a_0 + 0.1L_0^{(1)} + 0.1L_0^{(2)} - 0.4L_0^{(3)} + 0.5L_1^{(1), a_0, c_0 = 0} - 0.5L_0^{(5)}.
\end{align*}
The variance of $Y_1^{a_0, c_0 = 0}$ is 1, and the variance of $L_1^{(1), a_0, c_0 = 0}$ and $L_1^{(2), a_0, c_0 = 0}$ are both 0.4 and the off-diagonal elements of the variance matrix for $\bm{L}_1$ are all zero. 

The treatment at time one follows a Bernoulli distribution with parameter
\begin{align*}
    &\Pr \left[A_1^{a_0, c_0 = 0} = 1 | \bm{L}_0, \bm{L}_1^{a_0, c_0 = 0}\right]\\
    =& \text{expit}\left[-1 + 0.1L_0^{(1)} + 0.1L_0^{(2)} + 0.2L_0^{(3)} + 0.2L_0^{(4)} - L_1^{(1),a_0, c_0 = 0} - 0.5L_1^{(2), a_0, c_0 = 0}\right].
\end{align*}
The censoring indicator at time 1 follows a Bernoulli distribution with parameter
\begin{align*}
    &\Pr\left[C_1^{\bm{a}, c_0=0}=1| \bm{L}_0, \bm{L}_1^{a_0, c_0 = 0}\right]\\ 
    =& \text{expit}\left[-4 + 0.3a_0 + 0.5a_1 + 0.3L_0^{(1)} - 0.3L_0^{(2)} - 0.3L_0^{(3)} + 0.1L_1^{(1), a_0, c_0 = 0} + 0.1L_0^{(5)}\right].
\end{align*}
The outcome $Y_2^{\bm{a}, \bm{c} = \bm{0}}$ follows a normal distribution with mean 
\begin{align*}
     \E\left[Y_2^{\bm{a}, \bm{c} = \bm{0}} | \bm{L}_0, \bm{L}_1^{a_0, c_0 = 0}\right] =&  -2 + 0.1a_0 + 0.1a_1 + 0.3L_0^{(1)} - 2 a_0 I(L_0^{(2)} > 0.5) - 2 a_1 I(L_0^{(2)} > 0.5) \\
    & - 0.3L_0^{(3)} + 2L_1^{(1), a_0, c_0 = 0} + 2L_1^{(2), a_0, c_0 = 0}
\end{align*}
and variance 1. We simulated $12,000$ observations and randomly split the dataset into an initial tree building dataset consisting of 10,000 observations and a final tree selection dataset which consisted of the remaining 2,000 observations. We focused on contrasting treatment regimes $\bm{a}_1 = \bm{1} = (1, 1)$ and $\bm{a}_0 = \bm{0} = (0, 0)$. For this data generating distribution the treatment effect differs depending on whether $L_0^{(2)}>0.5$ or not and the correct tree splits on $L_0^{(2)}$ at split-point 0.5.

When estimating the models for the probability of treatment assignment, the probability of dropping out of the study, and the model for the expectation of the pseudo-outcome at each time, we used generalized linear regression models with linear and additive main effects for all past covariate and treatment variables. We used 1,000 simulations to evaluate the performance of the SDLD algorithm. 

To evaluate the algorithm we used the following evaluation measures:
\begin{itemize}
    \item Proportion of correct trees: We say a tree is correct if it splits the correct number of times on each variable \citep{steingrimsson2016doubly}. In our setup this means splitting once on $L_0^{(2)}$ and never on other variables.
    \item The  number of terminal nodes in the tree.
    \item The number of noise variables that are split on (i.e.,~variables which are not treatment effect modifiers).
    \item Proportion ot times the first split in the initial tree is on $L_0^{(2)}$.
    \item  Pairwise prediction similarity \citep{steingrimsson2016doubly}: Pairwise prediction similarity measures the ability of the SDLD algorithm to stratify observations into correct groups and is defined as $1-\sum_{i = 1}^{1000}\sum_{j>1}^{1000} \frac{\left|I_T(i, j) - I_M(i, j)\right|}{\binom{1000}{2}}$, where $I_T(i, j)$ and $I_M(i, j)$ are indicators for whether individuals $i$ and $j$ fall in the same terminal node running down the correct tree and the estimated tree, respectively. The pairwise prediction similarity is bounded between zero and one with higher values indicating better performance.
\end{itemize}

\subsection{Simulation Results}

The SDLD algorithm identified the correct tree in 89.5\% of the simulation runs and the pairwise prediction similarity was 0.98. The average size of the tree was 2.2, which is close to the size of the correct tree, 2. The average number of noise variables the tree splits on was 0.16. The initial tree built by the initial tree building step first splits on $L_0^{(2)}$ (i.e.,~the correct covariate we expect the tree to first split on) in all simulation runs.

\section{Additional information for the analysis of the AMPATH data} \label{app: anlss_ampth}

In this section we provide additional details on the analysis in Section \ref{sec: ampath_anlss}. For the analysis all continuous variables were standardized. 

\textit{Definition of treatment:} When defining treatment assignment in each time period, we focused on visits where either a) an individual was assigned to at least three medications \citep{world2016consolidated} or b) an individual was assigned to less than three medications and one of them was DTG. We defined treatment status as being on DTG ($A_k$=1) if the individual was on DTG at the last visit in the time period and not being on DTG  ($A_k$=0) if the individual was not on DTG at the last visit in the time period. Treatment assignment was considered missing if there was no visit within a time period with available treatment information that satisfies either conditions a) or b) listed above.  

\textit{Definition of outcome and time-varying covariates:} 
To define the values of time-varying covariates in each time period (including the value of the outcome) we used data from the last visit before or at the visit where treatment was captured if there was no switch of DTG status in the time period and before or at the last visit when a switch happened if there was switch of DTG status in the time period. If treatment was missing in the time period, we used the last observed value of time-varying covariates in the time period. In the last time period ($K+1$), we used the last observed outcome measure as $Y_{K+1}$. If there were no observed measurements satisfying the aforementioned condition, outcome and/or time-varying covariates were considered missing in the time period. 

If outcome or treatment information was not available at time $0$, we re-defined day 0 as the date of the first visit. If weight was still missing at time 0 after re-defining day 0, we further re-defined day 0 as the date of the visit with the first weight measurement. If treatment was still missing at time 0 after re-defining day 0, we move the date of visit with the first observed treatment to the last day of time 0 (i.e., day $d-1$ for a window length of $d$ days); therefore, day 0 was re-defined as $(d-1)$ days before date of visit with the first observed treatment. After time $0$ we used carry-over from the previous time periods to handle missing outcome and treatment information. We excluded participants who still had missing values in outcome and treatment at time 0 after being pre-processed as described above. That is, we excluded participants with 1) no observed weight measurements, 2) no observed treatment, or 3) no observed treatment after the first observed weight measurement.



To handle missingness in viral load, we created an ordinal variable with missingness as the lowest category. The new variable had four categories ordered as 1) missing, 2) $[0, 10^3)$, 3) $[10^3, 10^4)$, and 4) $[10^4, \infty)$. In the analysis, we dropped all covariates that had more than 10\% missing values. For the remaining time-varying covariates (most with $\leq 1.5\%$ missing data), we handled missing data by carrying over values from the previous time period if there were observed measurements to carry over; otherwise for continuous covariates, we imputed the missing values using the average of the observed values in the time period, and for binary covariates, we imputed the missing values by 0. Tables \ref{tab: cov_drop} and \ref{tab: cov_imp} show the number and proportion of missing values in covariates that were dropped and in time-varying covariates that were used, respectively. Table \ref{tab: cov_summ} present the summary for each covariate used in the analysis before and after imputation and the two distributions are very similar.


            \begin{table}[htbp]
                \centering
                \begin{tabular}{lcccc}
                \toprule
                    Covariate & Visit 0 (\%) & Visit 1 (\%)  & Visit 2 (\%) & Visit 3 (\%)\\
                    \quad Uncensored individuals & $n = 84,445$ & $n = 71,431$ & $n = 64,353$ & $n = 58,646$ \\
                    \midrule
                    Employed outside home & 31,807 (37.7\%) & - & - & -\\
                    Electricity in home & 31,669 (37.5\%) & - & - & -\\
                    Water piped in home & 31,928 (37.8\%) & - & - & -\\
                    CD4 count & 77,760 (92.1\%) & 66,453 (93.0\%) & 60,252 (93.6\%) & 55,306 (94.3\%) \\ 
  Blood glucose & 84,407 ($>99.9$\%) & 71,376 (99.9\%) & 64,300 (99.9\%) & 58,582 (99.9\%) \\ 
  Fasting serum glucose & 84,438 ($>99.9$\%) & 71,420 ($>99.9$\%) & 64,338 ($>99.9$\%) & 58,628 ($>99.9$\%) \\ 
                     \bottomrule
                \end{tabular}
                \caption{\label{tab: cov_drop} Number and proportion of missing values in covariates that were dropped from the analysis.}
            \end{table}

                         \begin{table}[htbp]
                \centering
                \begin{tabular}{@{}lcccccccccccccc@{}}
                \toprule
                    Covariate & Visit 0 & Visit 1 & Visit 2 & Visit 3\\
                     \quad Uncensored individuals & $n = 84,445$ & $n = 71,431$ & $n = 64,353$ & $n = 58,646$ \\
                    \midrule
                    Systolic blood pressure & 1,269 (1.5\%) & 179 (0.3\%) & 66 (0.1\%) & 35 (0.1\%) \\ 
  Diastolic blood pressure & 1,276 (1.5\%) & 181 (0.3\%) & 68 (0.1\%) & 36 (0.1\%) \\ 
  Height & 4,276 (5.1\%) & 1,659 (2.3\%) & 862 (1.3\%) & 405 (0.7\%) \\ 
  TB treatment & 56 (0.1\%) & 11 (0\%) & 4 (0\%) & 3 (0\%) \\ 
  Married/living with partner & 3,019 (3.6\%) & 484 (0.7\%) & 174 (0.3\%) & 71 (0.1\%) \\ 
  Covered by NHIF & 6,362 (7.5\%) & 2,305 (3.2\%) & 1,039 (1.6\%) & 496 (0.8\%) \\ 
                     \bottomrule
                \end{tabular}
                \caption{\label{tab: cov_imp} Number of proportion of missing values in time-varying covariates that were used in the analysis. TB refers to tuberculosis; NHIF refers to National Health Insurance Fund.}
            \end{table}
            
                        \begin{table}[htbp]
                \centering
                \begin{tabular}{@{}lccccccccccccc@{}}
                \toprule
                      & \multicolumn{2}{c}{$\bm{L}_0$} & & \multicolumn{2}{c}{$\bm{L}_1$} \\
                      & \multicolumn{2}{c}{$(n = 84,445)$} & & \multicolumn{2}{c}{$(n = 71,431)$} \\
                     \cmidrule[0.5pt]{2-3} \cmidrule[0.5pt]{5-6}
                      Covariates & Before & After &  & Before & After\\
                     \midrule
                     Male, $n (\%)$ & \multicolumn{2}{c}{28,440 (33.7\%)} & & \multicolumn{2}{c}{-}\\
                     Age when starting ART, years & \multicolumn{2}{c}{$38.1 \pm 10.4$} & & \multicolumn{2}{c}{-}\\ 
                     Age at visit 0, years & \multicolumn{2}{c}{$42.1 \pm 11.0$} & & \multicolumn{2}{c}{-} \\
                     Time on ART at time 0, years & \multicolumn{2}{c}{$4.0 \pm 3.9$} & & \multicolumn{2}{c}{-}\\
                     Whether enrolled on or  & \multicolumn{2}{c}{24,688 (29.2\%)} & & \multicolumn{2}{c}{-} \\
                     \qquad after July 1, 2016, $n (\%)$ \\
                     Weight, kg & \multicolumn{2}{c}{$62.2 \pm 12.5$} & & \multicolumn{2}{c}{-} \\
                     Viral load, $n (\%)$ \\
                     \qquad Missing & \multicolumn{2}{c}{48,309 (57.2\%)} & & \multicolumn{2}{c}{10,043 (14.1\%)}\\
                     \qquad $[0, 10^3)$ & \multicolumn{2}{c}{30,025 (35.6\%)} & & \multicolumn{2}{c}{54,954 (76.9\%)}\\
                     \qquad $[10^3, 10^4)$ & \multicolumn{2}{c}{3,166 (3.7\%)} & & \multicolumn{2}{c}{3,459 (4.8\%)}\\
                     \qquad $[10^4, \infty)$ & \multicolumn{2}{c}{2,945 (3.5\%)} & & \multicolumn{2}{c}{2,975 (4.2\%)} \\
                     Systolic blood pressure, mmHg & $118.9 \pm 18.7$ & $118.9 \pm 18.6$ & & $119.8 \pm 18.9$ & $119.8 \pm 18.9$\\
                     Diastolic blood pressure, mmHg & $73.9 \pm 12.1$ & $73.9 \pm 12.0$ & & $73.9 \pm 12.3$ & $73.9 \pm 12.2$\\
                     Height, cm & $165.7 \pm 9.2$ & $165.7 \pm 9.0$ & & $165.6 \pm 9.5$ & $165.6 \pm 9.3$\\
                     Active TB, $n (\%)$ & \multicolumn{2}{c}{2,821 (3.3\%)} &  & \multicolumn{2}{c}{2,479 (3.5\%)} \\
                     TB treatment, $n (\%)$ & 1,936 (2.3\%) & 1,936 (2.3\%) & & 596 (0.8\%) & 596 (0.8\%)\\
                     Married/living with partner, $n (\%)$ & 47,759 (58.7\%) & 47,759 (56.6\%) & & 41,065 (57.9\%) & 41,065 (57.5\%)\\
                     Covered by NHIF, $n (\%)$ & 16,485 (21.1\%) & 16,485 (19.5\%) & & 16,104 (23.3\%) & 16,104 (22.5\%)\\
                     \midrule
                      & \multicolumn{2}{c}{$\bm{L}_2$} & &  \multicolumn{2}{c}{$\bm{L}_3$}\\
                      & \multicolumn{2}{c}{$(n = 64,353)$} & &  \multicolumn{2}{c}{$(n = 58,646)$}\\
                      \cmidrule[0.5pt]{2-3} \cmidrule[0.5pt]{5-6}
                      & Before & After & & Before & After\\
                      \midrule
                     Viral load, $n (\%)$ & \\
                     \qquad Missing & \multicolumn{2}{c}{4,622 (7.2\%)} & & \multicolumn{2}{c}{2,368 (4.0\%)}\\
                     \qquad $[0, 10^3)$ & \multicolumn{2}{c}{54,117 (84.1\%)} & & \multicolumn{2}{c}{51,749 (88.2\%)}\\
                     \qquad $[10^3, 10^4)$ & \multicolumn{2}{c}{3,269 (5.1\%)} & & \multicolumn{2}{c}{2,797 (4.8\%)}\\
                     \qquad $[10^4, \infty)$ & \multicolumn{2}{c}{2,345 (3.6\%)} & & \multicolumn{2}{c}{1,732 (3.0\%)} \\
                     Systolic blood pressure, mmHg & $120.6 \pm 18.9$ & $120.6 \pm 18.9$ & & $121.2 \pm 18.8$ & $121.2 \pm 18.8$\\
                     Diastolic blood pressure, mmHg & $74.6 \pm 12.1$ & $74.6 \pm 12.1$ & & $75.1 \pm 12.1$ & $75.1 \pm 12.1$\\
                     Height, cm & $165.7 \pm 9.3$ & $165.7 \pm 9.2$ & & $165.8 \pm 9.0$ & $165.8 \pm 9.0$\\
                     Active TB, $n (\%)$ & \multicolumn{2}{c}{2,349 (3.7\%)} &  & \multicolumn{2}{c}{2,167 (3.7\%)} \\
                     TB treatment, $n (\%)$ & 369 (0.6\%) & 369 (0.6\%) & & 277 (0.5\%) & 277 (0.5\%)\\
                     Married/living with partner, $n (\%)$ & 37,061 (57.7\%) & 37,061 (57.6\%) & & 33,164 (56.6\%) & 33,164 (56.5\%)\\
                     Covered by NHIF, $n (\%)$ & 15,970 (25.2\%) & 15,970 (24.8\%) & & 15,779 (27.1\%) & 15,779 (26.9\%)\\
                     \bottomrule
                \end{tabular}
                \caption{Summary of covariates used in the analysis before and after imputation. For binary or ordinal covariates, we present the number and proportion of participants falling in the listed category; for continuous covariates, we present the mean and standard deviation. For covariates with no missing values, we present one summary for before and after imputation. ART refers to antiretroviral therapy; TB refers to tuberculosis; NHIF refers to National Health Insurance Fund.}
                \label{tab: cov_summ}
            \end{table}
            
            Table \ref{tab: avg_eff} shows the estimated average weight had all individuals been always and never on DTG-containing ARTs and the average weight gain when comparing always being on DTG-containing ARTs to never being on DTG-containing ARTs supporting Figure \ref{fig: avg_eff}. The confidence intervals are estimated using the non-parametric bootstrap with 10,000 bootstrap samples. 
                \begin{table}[htbp]
    \centering
    \begin{tabular}{ccccccccccc}
    \toprule
    Time & Uncensored & \multicolumn{2}{c}{Always DTG} & & \multicolumn{2}{c}{Never DTG}\\
    \cmidrule[0.5pt]{3-4} \cmidrule[0.5pt]{6-7}
          $k$ & $n$ & $n$ & $\widehat{E}\left[Y^{\overline{\bm{1}}_{k-1}, \overline{\bm{c}}_{k-1} = 0}_k\right]$ & & $n$ & $\widehat{E}\left[Y^{\overline{\bm{0}}_{k-1}, \overline{\bm{c}}_{k-1} = 0}_k\right]$ & DTG Effect \\
         \midrule
1 & 71,431 & 5,700 & 63.27 (63.05, 63.49) & & 65,731 & 62.18 (62.09, 62.27) & 1.09 (0.88, 1.30) \\ 
  2 & 64,353 & 3,083 & 63.81 (63.56, 64.06) & & 59,550 & 62.60 (62.51, 62.70) & 1.20 (0.96, 1.44) \\ 
  3 & 58,646 & 1,350 & 64.10 (63.79, 64.41) & & 53,164 & 62.83 (62.73, 62.92) & 1.28 (0.96, 1.59) \\ 
  4 & 54,005 & 517 & 64.59 (64.17, 65.00) & & 44,473 & 63.18 (63.08, 63.28) & 1.41 (0.99, 1.82) \\ 
         \bottomrule
    \end{tabular}
    \caption{\label{tab: avg_eff} The table shows results from analysis of the whole analysis set. The table shows the total number of individuals with data available at each time $k$ and the number of individuals that were always or never on DTG-containing ART up to that time. The column labeled $\widehat{E}\left[Y^{\overline{\bm{1}}_{k-1}, \overline{\bm{c}}_{k-1} = 0}_k\right]$ shows the estimated weight (kg) using targeted maximum likelihood estimation (TMLE) if always on DTG-containing ARTs and the column labeled $\widehat{E}\left[Y^{\overline{\bm{0}}_{k-1}, \overline{\bm{c}}_{k-1} = 0}_k\right]$ shows the estimated weight if never on DTG-containing ARTs. The column labeled DTG Effect shows the estimated causal effect of always vs. never being on a DTG-containing ART on weight gain since time 0.}
\end{table}

\begin{figure}
    \centering
    \includegraphics[width = \textwidth]{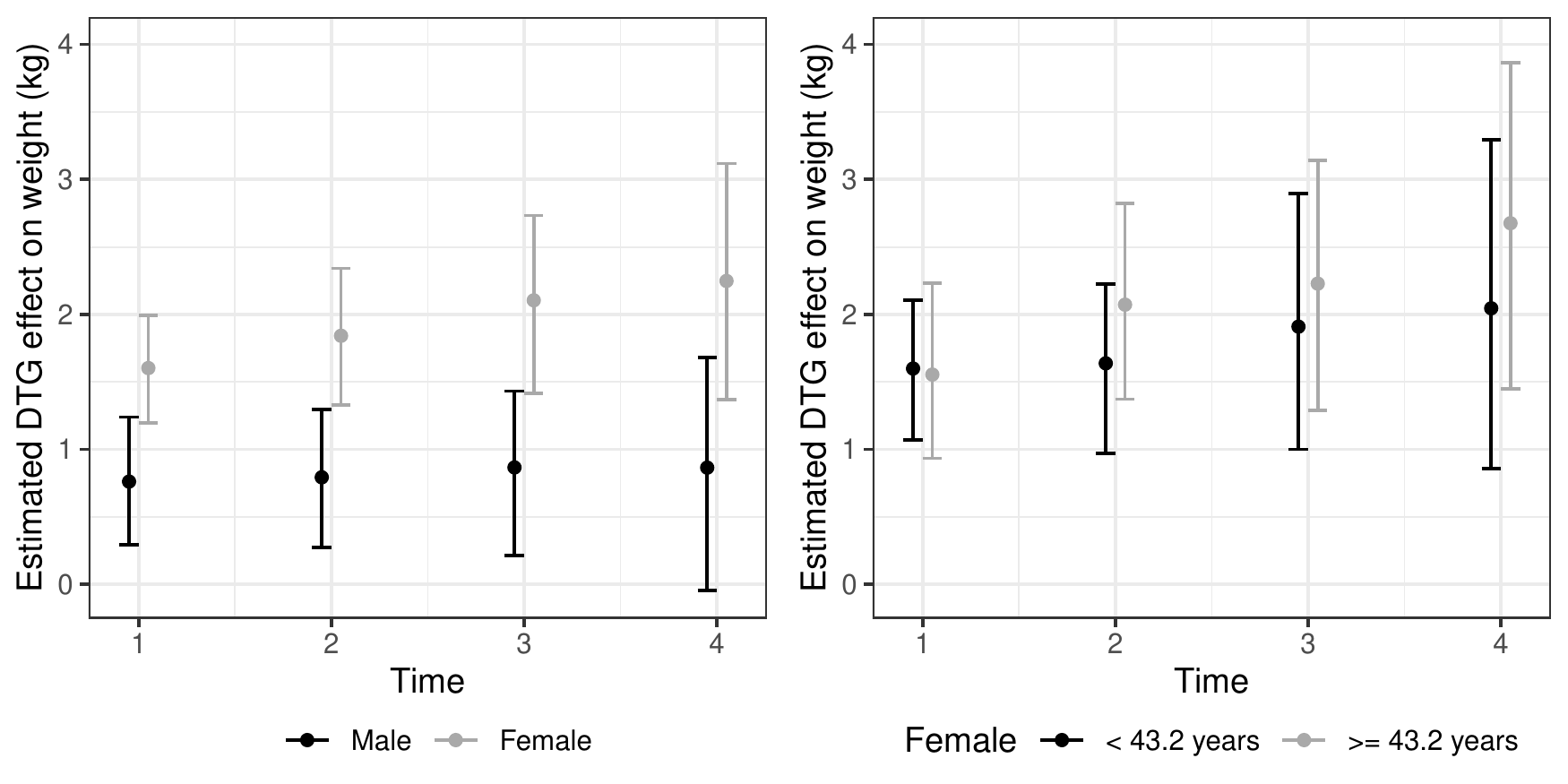}
    \caption{\label{fig: tree_trjctr} Estimated causal effect at each time on average weight in the subgroups discovered by the SDLD algorithm comparing always vs. never being on a DTG-containing ART. The left panel shows the average effect among males and females and the right panel shows the average effect among females who are younger and older than or equal to 42.8 years old.}
\end{figure}

Figure \ref{fig: tree_trjctr} plots the estimated average weight gain from receiving DTG-containing ART at each time for the subgroups discovered by the SDLD algorithm. The left panel shows the average effect among males and females and the right panel shows the average effect among females younger and older than or equal to 42.8 years old. The results show that both males and females gain most weight in the first 200 days; there is almost no additional weight gain after the first 200 days for males and there is a slight increase in weight gain from being on DTG-containing ART for a longer period of time among females. There is a large overlap of the confidence intervals for weight gain among younger and older females.

\textit{Stability of the SDLD algorithm:} Table \ref{tab: size_tree} presents the number of subgroups having differential effects of DTG-containing ARTs on weight gain discovered by the SDLD algorithms for 1,000 different data-splits. Table \ref{tab: splits} presents the number of times the final trees make splits on the variables in $\bm{L}_0$ across different splits of the data. 

\begin{table}[htbp]
    \centering
    \begin{tabular}{ccccccc}
    \toprule
         Subgroups & 1 & 2 & 3 & 4 \\
         \midrule
         $n$ & 245 & 206 & 322 & 227 \\
         \bottomrule
    \end{tabular}
    \caption{\label{tab: size_tree} Number of subgroups discovered by the SDLD algorithm across different data-splits}
\end{table}

            \begin{table}[htbp]
                \centering
                \begin{tabular}{lccccccc}
                \toprule
                     Covariate & First split & Other splits & Total \\
                     \midrule
                     Male & 482 &  61 & 543 \\ 
  Age when starting ART &  66 &  90 & 156 \\ 
  Age at time 0 &  64 & 264 & 328 \\ 
  Time on ART at time 0 &  13 &   2 &  15 \\ 
  Whether enrolled on or after July 1, 2016 &  11 &   0 &  11 \\ 
  Weight &  14 & 138 & 152 \\ 
  Viral load &   3 &  46 &  49 \\ 
  Systolic blood pressure &  12 &  58 &  70 \\ 
  Diastolic blood pressure &   9 &  34 &  43 \\ 
  Height  &  69 &  56 & 125 \\
  Active TB &   0 &   0 &   0 \\ 
  TB treatment &   0 &   0 &   0 \\ 
  Married/living with partner &  12 &  25 &  37 \\ 
  Covered by NHIF &   0 &   2 &   2 \\ 
                     \bottomrule
                \end{tabular}
                \caption{\label{tab: splits} Number of times the final tree  splits on each variables in $\bm{L}_0$ across the $1000$ different data splits. ART refers to antiretroviral therapy; TB refers to tuberculosis; NHIF refers to National Health Insurance Fund.}
                
            \end{table}

\end{document}